\documentclass[useAMS,usenatbib]{mn2e}

\usepackage{epsfig}
\usepackage{longtable}
\usepackage{times}

\usepackage{amsmath}
\usepackage{amssymb}
\def \inte {$INTEGRAL$}
\def \xmm {$XMM$-$Newton$}

\def \chandra {$Chandra$}

\def \src {CXOU~J191043.7+091629}
\def \ax {AX~J1910.7+0917}
\def \xmmname {2XMM~J191043.4+091629}

\def \nh {N${\rm _H}$}

\def \hcm {\hbox {\ifmmode $ atom cm$^{-2}\else atom cm$^{-2}$\fi}}
\def \arcmin {\hbox{$^\prime$}}
\def \arcsec {\hbox{$^{\prime\prime}$}}

\def \apj {ApJ}

\def \apjl {ApJL}
\def \apjs {ApJS}
\def \aap {A\&A}

\def \mnras {MNRAS}

\def \aapr {A\&A Rev.}

\def\xmm {\emph{XMM-Newton}}
\def\cxo {\emph{Chandra}}

\def\flux {\mbox{erg cm$^{-2}$ s$^{-1}$}}
\def\lum {\mbox{erg s$^{-1}$}}
\def\nh {$N_{\rm H}$}
\newcommand{\be}{\begin{equation}}

\newcommand{\ee}{\end{equation}}

\title[\ax: the slowest X--ray pulsar]{\ax: the slowest X--ray pulsar
}
\author[Sidoli et al.]{L.~Sidoli,$^{1}$\thanks{E-mail: sidoli@iasf-milano.inaf.it} G.~L.~Israel,$^{2}$ P.~Esposito,$^{3}$ G.~A.~Rodr\'iguez Castillo,$^{2}$ and K.~Postnov$^{4, 5}$ \\
$^{1}$INAF, Istituto di Astrofisica Spaziale e Fisica Cosmica, Via E.\ Bassini 15,   I-20133 Milano,  Italy   \\
$^{2}$INAF, Osservatorio Astronomico di Roma, Via Frascati 33, I-00040 Monteporzio Catone, Italy \\
$^{3}$Anton Pannekoek Institute for Astronomy, University of Amsterdam, Postbus 94249, NL-1090-GE Amsterdam, The Netherlands \\
$^{4}$ Moscow Lomonosov State University, Faculty of Physics, Leninskie Gory 1/2, 119991, Moscow Russia \\
$^{5}$ Moscow Lomonosov State University, Sternberg Astronomical Institute 13, Universitetskij pr., 119234,  Moscow, Russia \\
}

\begin{document}

\date{Accepted 2017 May 3. Received 2017 May 3; in original form 2017 April 6}

\pagerange{\pageref{firstpage}--\pageref{lastpage}} \pubyear{2017}

\maketitle

\label{firstpage}

\begin{abstract}
Pulsations from the high mass X-–ray binary \ax\ were
discovered during \cxo\ observations performed in 2011 \citep{Israel2016}.
We report here more details on this discovery and discuss the source nature.
The period of the X-ray signal is  $P = 36200\pm110$~s, with a pulsed fraction, $PF$, of 63$\pm$4\%.
Given the association with a massive B-type companion star, 
we  ascribe this long periodicity to the rotation of the neutron star,
making \ax\ the slowest known X--ray pulsar. 
We report also on the spectroscopy of \xmm\ observations that serendipitously covered the source field,
resulting in an highly absorbed (column density almost reaching $10^{23}$~cm$^{-2}$), power law X-ray spectrum. 
The X-ray flux is variable on a timescale of years, spanning a dynamic range $\gtrsim60$. 
The very long neutron star spin period can be explained within a quasi-spherical settling accretion model, 
that applies to low luminosity, wind-fed, X--ray pulsars.
\end{abstract}


        %%%%%%%%%%%%%%%%%%%%%%%%%%%%%%%%%%%%%%%%%%%%%%%%%%%%%%%%%
        \section{Introduction\label{intro}}
        %%%%%%%%%%%%%%%%%%%%%%%%%%%%%%%%%%%%%%%%%%%%%%%%%%%%%%%%%

\ax\  (also known as \xmmname) was discovered  during the 
ASCA Galactic Plane Survey and it was included in the catalogue of the faint point-like sources \citep{Sugizaki2001}.
Different X--ray missions serendipitously covered the source position, since
it lies at a projected distance of about 12\arcmin\ from the Supernova Remnant (SNR) W49,
target of many observations.

\citet{Pavan2011} performed a comprehensive investigation of all X--ray data that were publicly available 
at that time, analysing ASCA (performed in 1993), \xmm\ (2004) and \cxo\ observations (2008), finding  that \ax\ 
is a variable (maybe transient) X-ray source.
No periodicities in the X--ray light curves were found.
Their X--ray spectroscopy resulted in a highly absorbed power law spectrum, well described by a column density, \nh, of a few  $10^{22}$~cm$^{-2}$,
and a photon index ranging from  $\Gamma$=2.3$\pm{0.5}$ (at a 1--10 keV observed flux of $\sim$5$\times10^{-12}$~erg~cm$^{-2}$~s$^{-1}$; ASCA)
to  $\Gamma$=1.28$\pm{0.08}$ (at a 1--10 keV observed flux of 2.4$\times10^{-11}$~erg~cm$^{-2}$~s$^{-1}$; \xmm).
A faint emission line from iron at $\sim$6.4~keV was also evident in the \xmm\ 2004 spectra.
During the only available \cxo\ observation pointed on \ax\, the source was undetected, with a 
1--10~keV flux F$<$4$\times10^{-13}$~erg~cm$^{-2}$~s$^{-1}$ (1$\sigma$ upper limit; not corrected for the absorption; \citealt{Pavan2011}).

A refined X--ray position with respect to the ASCA survey could be estimated from \xmm/EPIC data (where the source
was caught at large off-axis angle; \citealt{Pavan2011}): R.A.$_{\rm J2000}$=19$^{\rm h}$10$^{\rm m}$43$\fs$39,  
Dec$_{\rm J2000}$=09${\degr}$16$\arcmin$30$\farcs$0, with an uncertainty of 2'' (90\% c.l.).
Using these sky coordinates, a 2MASS countperpart was at first associated with \ax\ \citep{Pavan2011}, 
favouring a high mass X-ray binary (HMXB) nature.
Later, this infrared (IR) source was resolved into two stars by \citet{Rodes-Roca2013} 
using the UKIDSS-GPS (United Kingdom Infrared Deep Sky Survey-Galactic Plane Survey; \citealt{Lucas2008}).
Only one of these two stars lies within the XMM error region. 
The near-infrared (NIR) spectrum of this counterpart, obtained with the 
Near-Infrared Camera Spectrometer (NICS) at the Telescopio Nazionale {\em Galileo} (TNG)
allowed them to identify it with a B-type star, confirming a massive binary nature for \ax\ \citep{Rodes-Roca2013}, although the
luminosity class (a supergiant or a Be star) could not be securely established.
However, these authors favoured a supergiant HMXB  located at a distance
d=16.0$\pm{0.5}$~kpc, following a comprehensive discussion of both X-ray and NIR properties.

\ax\ is also reported at hard energies (above 20 keV) as a variable HMXB in the 
\inte/IBIS source catalog \citep{Bird2016}.

During a systematic search for coherent signals in publicly available \cxo\ observations, a periodicity at 36.2~ks 
was found from a faint source (named \src) compatible with being \ax\ \citep{Israel2016}.

In this {\em paper},  we give details on this discovery and discuss the source nature.

 	 %%%%%%%%%%%%%%%%%%%%%%%%%%%%%%%%%%%%%%%%%%%%%%%%%%%%%%%%%%%%%%%%%%%%
 	 \section{Observations and Data Reduction}
         \label{data_redu}
  	 %%%%%%%%%%%%%%%%%%%%%%%%%%%%%%%%%%%%%%%%%%%%%%%%%%%%%%%%%%%%%%%%%%%%

The \src\ sky position was serendipitously observed  by \cxo\ and \xmm, during observations
pointed on the SNR W49. The results on the SNR were published by \citet{Lopez2013a} and \citet{Miceli2006}.
Results on some archival X--ray data of \ax\ were reported by \citet{Pavan2011},  
to which we refer the reader.
We have analysed  more recent \xmm\ and \cxo\ long exposures of the source field, summarized in Table~\ref{tab:log}.

%%%%%%%%%%%%%%%%%%%%%%%%%%%%%%%%%%%%%%%%%%%%%%%%%%%%%%%%%%%%%%%%%%%%%%%%%%%%%%%%%%%%%%%%%%%%%%
\begin{table*}
\begin{minipage}{16.cm}
\centering \caption{Summary of the X-ray observations.} 
\label{tab:log}
\begin{tabular}{@{}lccccc}
\hline
Satellite & Instrument & Obs.\,ID  & Date & Exp. & Mode$^{a}$\\
&  & & & (ks) & \\
\hline
\cxo & ACIS & 13440 & 2011 Aug 18 & 160.0 & TE FAINT (3.24\,s)\\
\cxo & ACIS & 13441 & 2011 Aug 21 & 60.5 & TE FAINT (3.24\,s)\\
\emph{XMM} & MOS\,1 / MOS\,2 & 0084100601 & 2004~Apr~13 & 2.5 / 2.5 & FF (2.6\,s)  / FF (2.6\,s)  \\
\emph{XMM} & pn / MOS\,2     & 0724270101 & 2014~Apr~17 & 115.3 / 116.8 & FF (73.4\,ms) / FF (2.6\,s)  \\
\emph{XMM} & pn / MOS\,2     & 0724270201 & 2014 Apr 19 & 67.4 / 69.8 & FF (73.4\,ms) / FF (2.6\,s) \\   
\hline
\end{tabular}
\begin{list}{}{}
\item[$^{a}$] TE: Timed Exposure, FAINT: Faint telemetry format, FF: Full Frame, the readout time is given in parentheses.
\end{list}
\end{minipage}
\end{table*}
%%%%%%%%%%%%%%%%%%%%%%%%%%%%%%%%%%%%%%%%%%%%%%%%%%%%%%%%%%%%%%%%%%%%%%%%%%%%%%%%%%%%%%%%%%%%%%%

%%
During the two long \cxo/Advanced CCD Imaging Spectrometer (ACIS; \citealt{Garmire2003})
 observations performed in 2011 the source was located in one of the two ACIS-I, front-illuminated CCDs, operated
in full-imaging timed-exposure mode (no gratings) and with a frame time of 3.24~s. 
New level 2 event files were generated with the \cxo\ Interactive Analysis of Observations (\textsc{ciao}) software 
version 4.8, and products extracted adopting standard procedures.
We refer the reader to \citet{Israel2016} for additional details on the data reduction not contained in this paper.

The source field was covered by \xmm/EPIC (European Photon Imaging Camera) detectors, two of which 
use MOS CCDs \citep{Turner2001} while the third uses pn CCDs \citep{Struder2001}.
Due to the loss of one EPIC-MOS CCD (caused by a micro-meteorite\footnote{see https://www.cosmos.esa.int/web/xmm-newton/mos1-ccd3}), 
the \ax\ off-axis position
was imaged in 2014 only by MOS\,2 and pn CCDs (Table~\ref{tab:log}), both operated in full frame mode.
The data were reprocessed using version 15 of the Science Analysis Software ({\sc sas})
with standard procedures. 
Background counts were obtained from regions offset from the source position on the same CCD, at a similar offaxis distance.
The background level was stable along the observations, except than during the 2014 \xmm\ pointings,
where a further filtering was applied to exclude time intervals containing background flares. 
This led to reduced net exposure times   
of 98.2~ks (pn) and 109.4~ks (MOS~2) for  the  0724270101 observation, 
and of 30.8~ks (pn) and 34.7~ks (MOS~2) for the 0724270201 one.
Appropriate response matrices were generated
using the {\sc sas}  tasks {\sc arfgen} and {\sc rmfgen}.

For both \xmm\ and \chandra\ spectra, all  uncertainties  are given at 90\% confidence level for
one interesting parameter. 
Spectra were grouped to have a minimum of 30 counts per bin.
When fitting two EPIC spectra from a single observation, we fitted them simultaneously, 
adopting normalization factors to account for uncertainties in instrumental responses.
In the spectral fitting we adopted the interstellar abundances of \citet{Wilms2000}  
and  photoelectric absorption cross-sections of \citet{Verner1996}, 
using the absorption model  {\sc TBnew} in {\sc xspec}.

Arrival times have been corrected to the Solar System barycenter before searching for periodicities.

  	%%%%%%%%%%%%%%%%%%%%%%%%%%%%%%%%%%%%%%%%%%%%%%%%%%%%%
  	\section{Analysis and Results\label{result}}
  	%%%%%%%%%%%%%%%%%%%%%%%%%%%%%%%%%%%%%%%%%%%%%%%%%%%%%

%%%%%%%%%%%%%%%%%%%%%%%%%%%%%%%%%%%%%%%%%%%%%%%%%%%%%%%%%%%%%%%%%%%%%%%%%%%%%%%%%%%%%%%%%%%%%%%%%%%%%%%%%%%%%%%%
\subsection{X--ray position}
\label{sect:pos}
%%%%%%%%%%%%%%%%%%%%%%%%%%%%%%%%%%%%%%%%%%%%%%%%%%%%%%%%%%%%%%%%%%%%%%%%%%%%%%%%%%%%%%%%%%%%%%%%%%%%%%%%%%%%%%%%

\begin{figure*}
\centering
\resizebox{\hsize}{!}{\includegraphics[angle=0]{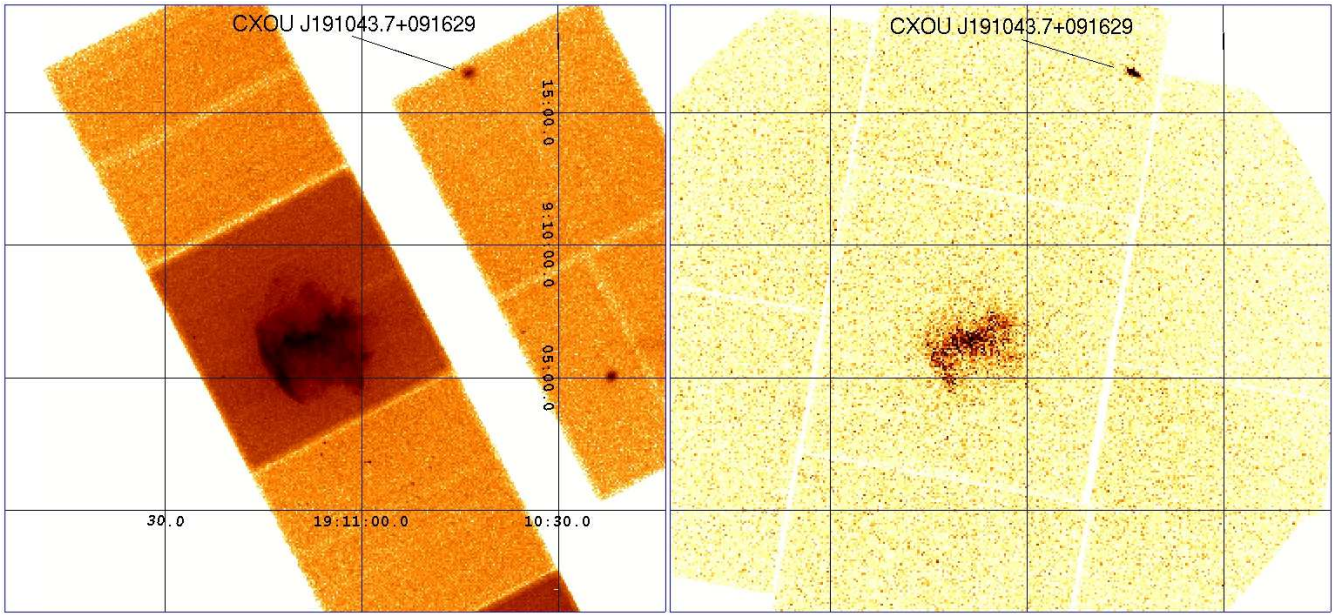}}
\caption{\src\ sky position serendipitously observed by \cxo\ (on the {\em left}, obsID 13441) 
and by \xmm\ (on the {\em right}, MOS~1, obsID 0084100601). The same scale applies to both images. The target was the SNR W49.
}
\label{fig:ima}
\end{figure*}

\ax\ is located at a very large off-axis position in the X-ray observations 
analysed in this paper (Fig.~\ref{fig:ima}), so 
unfortunately we are unable to provide refined X--ray sky coordinates, with respect to \citet{Pavan2011}.
The sky position we estimated from the
 \cxo\ observations is 
R.A.$_{\rm J2000}$=19$^{\rm h}$10$^{\rm m}$43$\fs$6,  
Dec$_{\rm J2000}$=09${\degr}$16$\arcmin$28$\farcs$9 (J2000), with an associated
1$\sigma$ error of 7\arcsec.
The big error region is due to the large off-axis source position \citep{Evans2010}. 
The \cxo\ centroid is $\sim$3.3\arcsec\ away from the XMM one \citep{Pavan2011}, and is 
compatible with the  infrared counterpart proposed
by \citet{Rodes-Roca2013}, an early-type massive companion.

%%%%%%%%%%%%%%%%%%%%%%%%%%%%%%%%%%%%%%%%%%%%%%%%%%%%%%

%%%%%%%%%%%%%%%%%%%%%%%%%%%%%%%%%%%%%%%%%%%%%%%%%%%%%%

%%%%%%%%%%%%%%%%%%%%%%%%%%%%%%%%%%%%%%%%%%%%%%%%%%%%%%%%%%%%%%%%%%%%%%%%%%%%%%%%%%%%%%%%%%%%%%%%%%%%%%%%%%%%%%%%
\subsection{Timing analysis}
\label{sect:pulsation}
%%%%%%%%%%%%%%%%%%%%%%%%%%%%%%%%%%%%%%%%%%%%%%%%%%%%%%%%%%%%%%%%%%%%%%%%%%%%%%%%%%%%%%%%%%%%%%%%%%%%%%%%%%%%%%%%

\begin{figure*}
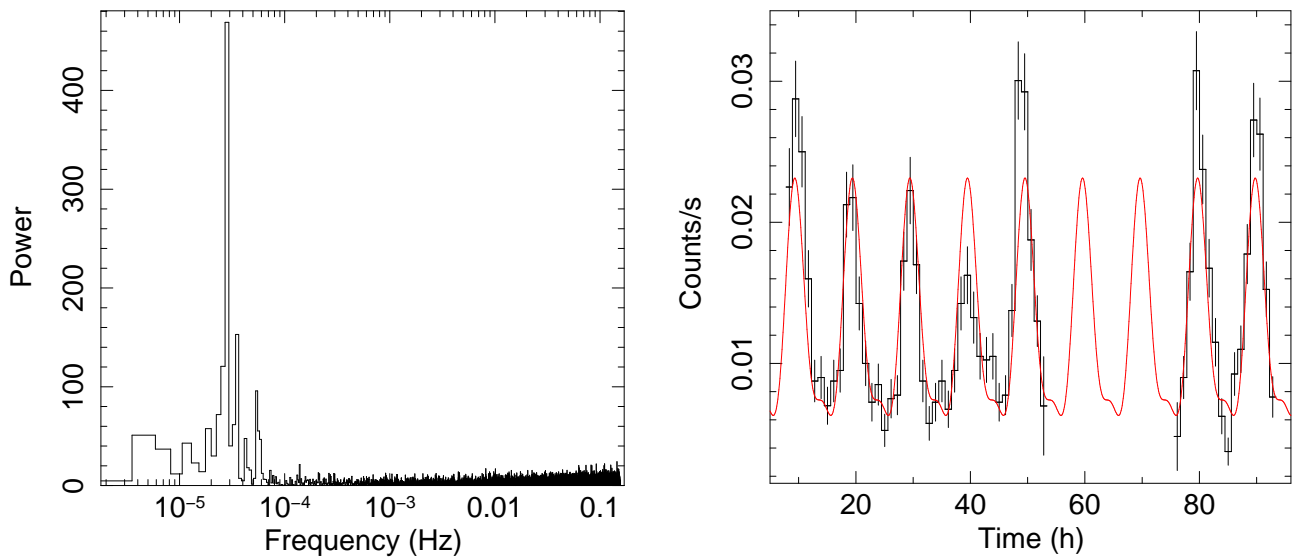

\begin{tabular}{cc}
\hspace{-0.5cm}
\includegraphics[height=8.9cm, angle=-90]{./C1910PSD.ps} 
\hspace{-0.2cm}
\includegraphics[height=8.9cm, angle=-90]{./C1910LC.ps} \\
\end{tabular}
\caption{Power density spectrum (PDS) obtained from the \cxo/ACIS data ({\em Left panel}) together with the  \cxo\ source light curve 
fitted with a sinusoidal model ({\em Right panel}).
}
\label{fig:pulsation}
\end{figure*}

As part of the \emph{Chandra ACIS Timing Survey} (CATS; \citealt{Israel2016}) 
project,\footnote{The CATS project is a Fourier-transform-based systematic and 
automatic search for new pulsating sources in the \cxo\ ACIS public archive, 
yielding, to date, up to 41 previously unknown X-ray pulsators.}
a coherent signal with a period of 36.2~ks was discovered from \ax.
The Fourier periodogram (normalized according to \citealt{Leahy1983}) 
is shown in Fig.~\ref{fig:pulsation} (left panel) together with the net light curve (right panel) fitted with a sinusoidal model. 
The periodic modulation is clearly evident, as well as a variability of the peak intensity at each cycle.
Adopting a phase-fitting technique \citep{Osso2003} using three time intervals, we estimated 
a final value for the periodicity of $P = 36200\pm110$~s (1$\sigma$), obtaining 
a pulsed fraction (semi-amplitude of the sinusoid divided by the source  average count rate) 
of $PF$=63$\pm$4\% (1$\sigma$).

%%%%%%%%%%%%%%%%%%%%%%%%%%%%%%%%%%%%%%%%%%%%%%%%%%%%%%%%
%%%%%%%%%%%%%%%%%%%%%%%%%%%%%%%%%%%%%%%%%%%%%%%%%%%%%%%%

%%%%%%%%%%%%%%%% 
The timing analysis of the \emph{XMM-Newton} data did not provide any
additional significant information about the periodic pulsation. The
inspection of the longest observation (0724270101) revealed a possible
modulation at $\sim$9.5~h. However, while the period is consistent
with the \emph{Chandra} one ($P_{\mathrm{XMM}}=34200\pm2200$~s at
3$\sigma$), the signal is not statistically significant (we estimate a
significance of $\approx$3$\sigma$); in this respect, we also note
that in the \emph{XMM-Newton} data set the upper limit on the pulsed
fraction for a modulation around 36~ks is larger than 100\%\ 
(as inferred from the PDS within a narrow frequency interval encompassing frequencies around
the \cxo\ period) and
therefore a non-detection would not be surprising. The successive
\emph{XMM-Newton} observation (0724270201) covered less than two
modulation cycles and was affected by episodes of strong flaring
background that hampered a proper search for periodicities.

\subsection{Spectroscopy}
\label{sect:spectra}

We performed the spectroscopy of \cxo\ and \xmm\ data adopting a simple, highly absorbed power law,
which already resulted in a good deconvolution of the X--ray emission.
The time-averaged spectral results for the five observations are reported in Table~\ref{tab:spec}.
While the absorbing column density is consistent with a constant value, the spectrum becomes harder when the source intensity is brighter.
The source X--ray flux spans a dynamic range of $\sim$60. 
The luminosity, calculated assuming a distance of 16 kpc \citep{Rodes-Roca2013}, ranges from 1.7$\times$$10^{34}$~\lum
 to $10^{36}$~\lum. The uncertainty on the X--ray luminosity in Table~\ref{tab:spec} includes only the error on
 the power law normalization.

Alternatively, a single absorbed hot blackbody model results in  an equally good fit to the \cxo\ and 2014 \xmm\ spectra,
implying a lower column density, \nh\, in the range 2-6$\times$$10^{22}$~cm$^{-2}$,  a temperature kT$_{\rm BB}$$\sim$1.1-1.9~keV,
and blackbody radii in the range R$_{\rm BB}$$\sim$100-300~m (at 16 kpc).
More complicated models are not required by the data.

\begin{figure*}
\centering
\resizebox{\hsize}{!}{\includegraphics[angle=-90]{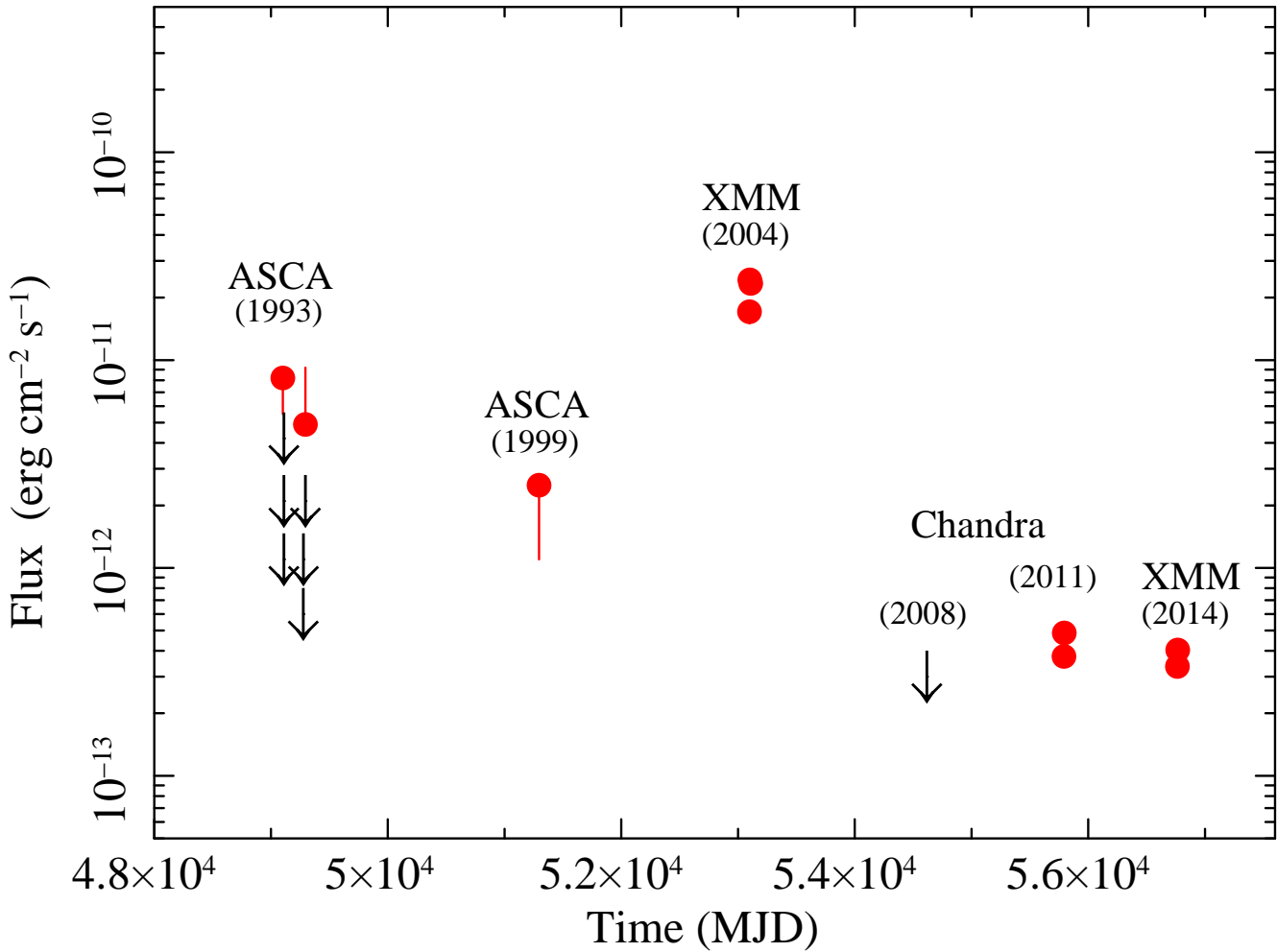}}
\caption{\ax\ long-term  light curve. Observed (not corrected for the absorption) X--ray fluxes in the energy range 1--10 keV have been collected
from \citealt{Pavan2011} and from the {\em present work}. When the error bars are not evident, they are smaller than the symbols.
}
\label{fig:history}
\end{figure*}

In Fig.~\ref{fig:history} we show the long-term \ax\ light curve, plotting the observed X--ray flux 
(not corrected for the absorption), in the energy range 1--10 keV.
We rely on \citet{Pavan2011} for \ax\ X--ray fluxes before 2004. All fluxes assume a power law continuum.

%%%%%%%%%%%%%%%%% 
\begin{table*}
\begin{minipage}{17.5cm}
\centering \caption{\cxo\ and \xmm\  spectroscopy with an absorbed power law model. }
\label{tab:spec}
\begin{tabular}{@{}lcccccc}
\hline
Obs.                  &  \nh\                 & $\Gamma$                   &  Unabs. flux (1-10 keV)     &    L$_{X}$ (1-10 keV)            & $\chi^2_\nu$ / dof \\
                      & ($10^{22}$~cm$^{-2}$) &                            & ($10^{-12}$~\flux)          &    ($10^{34}$~\lum)              &                     \\
%---------------------------------------------------------------------------------------------------------------------------------------------------------------------------------
\hline
\cxo/13440            &  $10^{+2}_{-1}$         & $2.3\pm{0.3}$              & $0.99\pm0.05$               &  $3.0\pm0.1$                      & 0.814 / 109          \\
\cxo/13441            &  $9\pm{2}$              & $2.0^{+0.5}_{-0.4}$        & $1.07\pm0.07$               &  $3.3\pm0.2$                      & 0.884 / 51         \\
%---------------------------------------------------------------------------------------------------------------------------------------------------------------------------------

\emph{XMM}/0084100601 & $7^{+3}_{-2}$        &  $0.9^{+0.5}_{-0.4}$        &  $33\pm{2}$                 &    $100\pm7$                     & 1.246 / 47            \\
\emph{XMM}/0724270101 & $8\pm{2}$            &  $1.5^{+0.3}_{-0.3}$        &  $0.57\pm{0.03}$            &    $1.7\pm0.1$                   & 1.114 / 58            \\
\emph{XMM}/0724270201 & $8^{+4}_{-3}$        &  $1.5^{+0.6}_{-0.5}$        &  $0.69\pm{0.08}$            &    $2.1\pm0.2$                   & 0.978 / 20             \\
%---------------------------------------------------------------------------------------------------------------------------------------------------------------------------------
\hline
\end{tabular}
\end{minipage}
\end{table*}

%%%%%%%%%%%%%%%%%%%%%%%%%%%%%%%%%%%%%%%%%%%%%%%%%%%%%%

%%%%%%%%%%%%%%%%%%%%%%%%%%%%%%%%%%%%%%%%%%%%%%%%%%%%%%

%%%%%%%%%%%%%%%%%%%%%%%%%%%%%%%%%%%%%%%%%%%%%%%%%%%%%%%%%
\section{Discussion}
%%%%%%%%%%%%%%%%%%%%%%%%%%%%%%%%%%%%%%%%%%%%%%%%%%%%%%%%%

A coherent signal at 36.2~ks from the source 
\src\ (\ax) was discovered during long \cxo\ observations performed in 2011 \citep{Israel2016}.
We have reported here details on this discovery, together with X--ray spectroscopy
of more recent \xmm\ pointings that serendipitously covered the source field.

\ax\ is a faint X-ray source the nature of which was previously investigated in depth by 
\citet{Pavan2011} and by \citet{Rodes-Roca2013}.
The latter authors performed a search for NIR counterparts, pinpointing
a single early-B massive star within the \xmm\ error circle provided by  \citet{Pavan2011}.
Although  \citet{Rodes-Roca2013} could not constrain the luminosity class of the B-type star,
being consistent with both a supergiant and a main sequence, Be star, they
favoured a supergiant HMXB located at a distance of 16.0$\pm{0.5}$~kpc, in the Outer Arm of our Galaxy. 
Since \ax\ was always detected at very large offaxis angles also in the more recent X-ray data-set analysed in our work,
we are unable to provide better sky coordinates (Sect.~\ref{sect:pos}) than the one already reported by \citet{Pavan2011}.
So we rely on these two previous papers for the source nature, where compelling evidence of an HMXB nature was found.

The X--ray properties we observed from the source in our data-set are  consistent with a HMXB nature as well:
the long-term X--ray light curve (Fig.~\ref{fig:history}) displays a dynamic range larger than 60, with an 
X-ray luminosity (at a distance of 16 kpc, \citet{Rodes-Roca2013}), ranging from 1.7$\times$$10^{34}$~\lum to $10^{36}$~\lum;
the X--ray emission is well charecterized by a highly absorbed power-law model, harder (in the range 1-10 keV) when the source is brighter, 
in agreement with what usually found in HMXB pulsars (see \citealt{Walter2015} and \citealt{Martinez-Nunez2017},
for the most recent reviews on HMXBs). 

Given the well-ascertained HMXB nature, the only viable interpretation of the $\sim$10~hr coherent signal 
is that it is the rotational period of the neutron star, making 
\ax\ the slowest X--ray pulsar known to date.
Indeed, before \ax, 
the slowest X--ray pulsar in an HMXB was 2S~0114+650 (associated with a B1Ia donor; \citealt{Grundstrom2007}), 
with a periodicity of $\sim$2.7~hr \citep{Finley1992, Hall2000, Hu2016}. 
A slowly rotating neutron star with a period of $\simeq 5.3$~hr
was also detected in a symbiotic X--ray binary with red giant donor 3A~1954+319
\citep{Marcu2011}, while an isolated pulsar with a very long spin period of 6.67~hr was discovered
within the supernova remnant RCW~103  \citep{DeLuca2006}.

\subsection{Quasi-spherical settling accretion}

The very slow period of the observed X-ray flux can be the spin period of a neutron star (NS) at the stage 
of quasi-spherical settling accretion in a wind-fed HMXB. 
This stage of accretion onto magnetized NS can be established when the X-ray luminosity of the source falls below some 
critical value $L_x\lesssim 4\times 10^{36}$~erg s$^{-1}$ (\citealt{Shakura2012}; see also \citealt{Shakura2017} for a recent update).  
In this regime, the accreting matter entry rate into NS magnetosphere, 
which determines the X-ray luminosity from the NS ($L_x\simeq 0.1 \dot M c^2$), is controlled by 
the plasma cooling due to Compton processes (at luminosities roughly above $10^{35}$~erg~s$^{-1}$) 
or radiative plasma cooling at lower luminosities.
 In a binary system with the orbital period $P_b=2\pi/\omega_B$, the torques acting on a magnetized NS rotating
with the period $P^*=2\pi/\omega^*$ read:
\begin{equation}
\label{susd}
I\dot\omega^*=Z\dot M\omega_BR_B^2-Z(1-z/Z)\dot M R_A^2\omega^*\,,
\end{equation}
where $I=10^{45}$~g~cm$^2$ is the NS moment of inertia, $Z>1$ is the dimensionless coupling coefficient between plasma and the 
NS magnetosphere (determined by the plasma cooling efficiency), $z\lesssim 1$ is the dimensionless factor describing the specific angular momentum of matter 
brought to the NS surface by accreting matter, $R_B=2GM_x/v_w^2$ is 
the gravitational capture (Bondi) radius of the NS from the stellar wind with the 
relative velocity $v_w$, $M_x$ is the NS mass, and $R_A$ is the Alfv\'en magnetospheric radius, 
which is determined by the pressure balance of matter near the magnetopsheric boundary. 
This equation can be conveniently recast to the relaxation form:
\begin{equation}
\label{rel}
\dot\omega^*=\frac{1}{\tau}(\omega^*_{eq}-\omega^*)
\end{equation}
where $\omega*_{eq}=(1-z/Z)^{-1}\omega_B(R_B/R_A)^2$ is the equilibrium NS frequency found from 
the vanishing torque condition $\dot\omega^*=0$, and $\tau$ is the characteristic time of reaching the equilibrium:
\begin{equation}
\label{tau}
\tau=\frac{I}{Z\dot M (1-z/Z) R_A^2}\,.
\end{equation}
For the characteristic values $\dot M=10^{14}$~g~s$^{-1}$ (corresponding to the X-ray luminosity $L_x=10^{34}$~erg~s$^{-1}$) and 
the canonical NS surface magnetic  field $B=10^{12}$~G 
(corresponding to the NS dipole magnetic moment $\mu=10^{30}\hbox{G\,cm}^3\mu_{30}$), 
we find (see \citealt{Shakura2017} for more detail and derivations)
that 
the magnetospheric radius is $R_A\sim 3\times 10^9$~cm, $Z\sim 5-10$ and $\tau\simeq 10^4$~years, 
the latter being only weakly dependent on the mass accretion rate $\dot M$. 
The solution of Eq. \ref{rel} reads: $\omega^*(t)=\omega^*_0e^{-t/\tau}+\omega^*_{eq}(1-e^{-t/\tau})$, 
where $\omega^*_0$ is the initial NS spin frequency. 
Even if $\omega^*_0\gg \omega^*_{eq}$, the NS frequency $\omega^*(t)$
tends to $\omega^*_{eq}$ in several $\tau$ intervals. 

The short relaxation time $\tau\sim 10^4$~years (keeping all the external parameters constant) suggests that 
in this accretion regime the NS rotation rapidly reaches the equilibrium  period $P_{eq}^*=2\pi/\omega^*_{eq}$:
\begin{equation}
\label{Peq}
P_{eq}\approx 5340\hbox{\,s\,}\mu_{30}^{12/11}\left(\frac{P_b}{10\hbox{\,d\,}}\right)\dot M_{14}^{-4/11}\left(\frac{v_w}{1000\hbox{\,km\,s}^{-1}}\right)^4\,.
\end{equation}
This formula immediately suggests that to observe $P_{eq}\approx 36200$~s, if we assume the standard NS magnetic field and the typical 
stellar wind velocity about 1000 km~s$^{-1}$, the binary system should have 
an orbital period of about 68 days, which is longer than typical orbital periods 
of supergiant HMXBs (usually, $P_b$$\sim$10~days, \citet{Liu2006}, but note that a large scatter is present 
and some outliers exist, 
like XTE~J1739-302,  with $P_b$=51~days, \citet{Drave2010}, and IGRJ~11215-5952, with $P_b$=165~days, \citet{Sidoli2007}). However,
increasing the NS magnetic field would decrease the orbital period estimate, $P_b\propto \mu^{-12/11}$, and 
moderate increase in the NS magnetic field by a few times is already sufficient to explain the observed 36.2 ks 
spin period of the X-ray pulsar in AX~J1910.7+0917 for more typical orbital periods of HMXBs with supergiants.

It is also easy to check that the mean X-ray luminosity $10^{34}-10^{35}$~erg~s$^{-1}$ can be obtained 
 assuming the typical values of the optical component mass $10 M_\odot$ and using the straightforward
 estimate of the captured wind rate by the NS in a HMXB [see Eq. (4) in \citealt{Shakura2014}]. 
Note, however, 
variable properties of the stellar wind can render settling accretion onto NS
highly unstable, with 
sudden flares and outbursts of different amplitudes, as discussed
for example in \citet{Shakura2014} in the context of Supergiant Fast X--ray Transient (SFXT) sources. 
It would be interesting to monitor the long-term X--ray flux variability of this source which may share some of SFXT properties. 

Therefore, the quasi-spherical settling accretion model for AX~J1910.7+0917 naturally provides explanation to the
observed record long spin period of accreting magnetized NS. Note here that in this regime 
in the limit of vanishing accretion rate 
the NS still continues spinnig-down by transferring angular momentum through the magnetospheric plasma shell, to
eventually reach the orbital period of the binary system, $P_{eq}^*|_{\dot M\to 0}\to P_b$. 
While technically challenging, even longer NS spin periods can be found in very faint Galactic X-ray binaries
in future sensitive X--ray observations.

%%%%%%%%%%%%%%%%%%%%%%%%%%%%%%%%%%%%%%%%%%%%%%%%%%%%%%%%%
\section{Conclusions}
%%%%%%%%%%%%%%%%%%%%%%%%%%%%%%%%%%%%%%%%%%%%%%%%%%%%%%%%%

A periodicity was discovered in \ax\ during \cxo\
observations that serendipitously covered the source region 
(\citealt{Israel2016}, see also Section~\ref{sect:pulsation}).
Since \ax\ is identified with a HMXB hosting a massive B-type star, 
the coherent signal at 36.2 ks can be explained only as the rotational period of the NS.
This discovery makes \ax\ the pulsar with the slowest spin period.
A quasi-spherical settling accretion model \citep{Shakura2012} is able to explain this superslow pulsation,
even adopting a typical NS surface magnetic field of $\sim$10$^{12}$~G. 
However, before a firm conclusion
about the NS magnetic field can be drawn from the theory, other source parameters (orbital period, velocity of the wind outflowing from the
massive donor) are needed.
Future sensitive spectroscopy above 10~keV would also be desirable, to detect cyclotron resonant scattering features
and obtain a direct measurement of the NS magnetic field in \ax.

%%%%%%%%%%%%%%%%%%%%%%%%%%%%%%%%%%%%%%%%%%%%%%%%%%%%%%%%%
\section*{Acknowledgments}
%%%%%%%%%%%%%%%%%%%%%%%%%%%%%%%%%%%%%%%%%%%%%%%%%%%%%%%%%

%----------
This work is based on data from observations with \xmm\ and \cxo.
\xmm\ is an ESA science mission with instruments and
contributions directly funded by ESA Member States and the USA (NASA).
The scientific results reported in this article are also based on data obtained 
from the \cxo\ Data Archive. This research has made use of 
software provided by the \cxo\ X-ray Center (CXC) in the application package \textsc{ciao}, 
and of the \textsc{simbad} database, operated at CDS, Strasbourg, France.
LS acknowledges the grant from PRIN-INAF 2014,  ``Towards a unified picture of accretion in High Mass X-Ray Binaries''  (PI: L.~Sidoli).
LS and KP acknowledge support from the International Space Science Institute (Switzerland) during 
a team meeting held in Bern in 2017 (PI: S.~Mart{\'{\i}}nez-N{\'u}{\~n}ez). 
KP also acknowledges partial support from RSF grant 16-12-10519 for travel to the ISSI meeting.
PE acknowledges funding in the framework of the NWO Vidi award A.2320.0076 (PI: N.~Rea).
\bibliographystyle{mn2e} 
\bibliographystyle{mnras}
%\bibliography{biblio}

\bsp

\label{lastpage}

\end{document}